# Study of deteriorating semiopaque turquoise lead-potassium glass beads at different stages of corrosion using micro-FTIR spectroscopy

Irina F. Kadikova[1, a)], Ekaterina A. Morozova[1, 2], Tatyana V. Yuryeva[1], Irina A. Grigorieva[3] and Vladimir A. Yuryev[4, b)]

[1] *The State Research Institute for Restoration, Building 1, 44 Gastello Street, Moscow 107114, Russia*
[2] *N. S. Kurnakov Institute of General and Inorganic Chemistry of the Russian Academy of Sciences, 31 Leninsky Avenue, Moscow, 119071, Russia.*
[3] *The State Hermitage Museum, 34 Dvortsovaya Embankment, Saint Petersburg, 190000, Russia.*
[4] *A. M. Prokhorov General Physics Institute of the Russian Academy of Sciences, 38 Vavilov Street, Moscow, 119991, Russia.*

a) E-mail: kadikovaif@gosniir.ru;
b) Corresponding author: Vladimir A. Yuryev, e-mail: vyuryev@kapella.gpi.ru

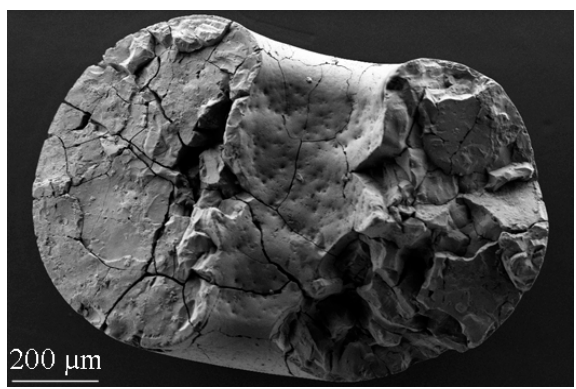

# Study of deteriorating semiopaque turquoise lead-potassium glass beads at different stages of corrosion using micro-FTIR spectroscopy

Nowadays, a problem of historical beadworks conservation in museum collections is actual more than ever because of fatal corrosion of the 19$^{th}$ century glass beads. Vibrational spectroscopy is a powerful method for investigation of glass, namely, of correlation of the structure-chemical composition. Therefore, Fourier-transform infrared spectroscopy was used for examination of degradation processes in cloudy turquoise glass beads, which in contrast to other color ones deteriorate especially strongly. Micro-X-ray fluorescence spectrometry of samples has shown that lead-potassium glass $PbO–K_2O–SiO_2$ with small amount of Cu and Sb was used for manufacture of cloudy turquoise beads. Fourier-transform infrared spectroscopy study of the beads at different stages of glass corrosion was carried out in the range from 200 to 4000 cm$^{-1}$ in the attenuated total reflection (ATR) mode. In all the spectra, we have observed shifts of two major absorption bands to low-frequency range (~1000 and ~775 cm$^{-1}$) compared to ones typical for amorphous $SiO_2$ (~1100 and 800 cm$^{-1}$, respectively). Such an effect is connected with $Pb^{2+}$ and $K^+$ appending to the glass network. The presence of a weak band at ~1630 cm$^{-1}$ in all the spectra is attributed to the adsorption of $H_2O$. After annealing of the beads, the band disappeared completely in less deteriorated samples and became significantly weaker in more destroyed ones. Based on that we conclude that there is adsorbed molecular water on the beads. However, products of corrosion (e.g., alkali in the form of white crystals or droplets of liquid alkali) were not observed on the glass surface. We have also observed glass depolymerisation in the strongly degraded beads, which is exhibited in domination of the band peaked at ~1000 cm$^{-1}$.

Keywords: nineteenth century glass; lead-potassium glass; translucent turquoise glass; seed beads; glass corrosion; ATR FTIR spectroscopy

**Introduction**

The appearance of glass beads was inseparably associated with origin of glass in the third millennium BC and further development of its manufacturing technology (Kachalov 1959, Dubin 2009). A short time later, Asia and Egypt—the main centres of glassmaking—began using opaque glass for making different articles including adornments. Soon, such things had become an important attribute in different rituals and ceremonies and played a huge role in life and culture of many countries and nations, as they represented power and wealth of their owners. In addition, such eminent works of art were essential objects for global trade expansion and were spread throughout the world before long.

Therefore, the glass beads items had been the important part of humanity everyday life since the early ancient times, but only in the mid of the 18$^{th}$ century the real flourishing of beadwork art began. Although, this period was rather short and lasted only until the end of the 19$^{th}$ century, a lot of artworks were created. In this period, the craft was extremely fashionable and popular in Europe, where the main centres of glassmaking manufacture were concentrated in Murano (Italy) and Bohemia (in the present-day Czech Republic) (Opper and Opper 1991). In addition, it was very important in North America, where American Indians used glass beads to decorate their clothing, religious objects, and functional tools and created unique style of work with glass beads (Lovell 2006). In Russia, beaded embroidery was especially loved throughout the whole 19$^{th}$ century. Both professional embroiderers and women of all walks of life created beaded items. Such enthusiasm was connected with several reasons. Firstly, development and the beginning of industrial-scale production of glass seed beads in the glass-making centres leaded to decreasing the cost. Then, the total number of varieties of glass beads, differing in size, colour and shape (round, faceted, multi-layered, transparent, cloudy, opaque, opalescent, etc.), exceeded a thousand (Yurova 2003). As a rule, 20 to 30 varieties of glass beads were used for embroidering a beadwork.

That is why the majority of museums in Europe, North America and Russia possess numerous exhibits, fully or partially comprised of beadworks, for instance, decorations, reticules (Figure 1a, b), different household items and even beaded pictures, as large as a few meters, with different plots (Yuryeva et al. 2018b).

Nowadays, the museum curators, conservators and restorers faced with a problem of preservation of historical beadworks from museum collections. It is an actual issue because of pernicious corrosion of the 19$^{th}$ century glass beads. There are many reports in the literature suggesting the glass corrosion, glass disease, or crizzling, i.e. chemical processes occurring on the surface of glass at high humidity, as the main cause of the destruction of glass (Brill 1975; Kunicki-Goldfinger 2008; O'Hern and McHugh 2014). Koob (2012) distinguishes five main stages of the corrosion process. At the first stage, hydrated alkali appears on the glass surface and glass becomes hazy. A characteristic feature of this stage is formation of droplets of liquid alkali on the glass surface (at high room humidity, usually above 55% RH) or alkali in the form of white crystals (at low humidity, usually under 40%). Then glass begins to crack and eventually this process becomes more and more aggravated, cracking goes deep into glass until the beads completely break up into separate fragments (the fifth stage of degradation process). However, restorers noticed that the one kind of beads—translucent turquoise lead-potassium ones—is subjected to destruction more intensively. Significantly, on the same sites of the embroidery both damaged and well-preserved turquoise beads are observed (Figure 1c). Therefore, we suggest that often not only crizzling, but mostly other processes (Figure 2) affect their appearance and state (Yuryeva and Yuryev 2014).

In the previous works, we established that formation of micro and nanocrystallites of orthorhombic $KSbOSiO_4$ (KSS) in glass matrix (Figure 2e–g), which arise during glass melting and bubbling, is the main cause of deteriorating of turquoise glass seed beads (Yuryeva et al. 2017, 2018a; Kadikova et al 2018). We concluded that KSS precipitates and their clusters give rise to internal stress (likely, tension) causing strain (likely, tensile one) in the glass bulk. In course of time it leads to the conspicuous 'internal corrosion' of glass resulted from its slow fracturing and gradual formation of a network of cracks (Yuryeva et al. 2017, 2018a, 2018b).

Another reason of degradation of translucent turquoise beads also relates to the process of their manufacturing (Yuryeva et al. 2018a, 2018b; Kadikova et al. 2018).

Recently, we have discovered that seed beads of many kinds are composed of two regions of different glass (Yuryeva et al. 2018a, 2018b). We have observed an internal core of the initial glass, often cracked due to KSS precipitates, and a crust coating the core (a shell) and consisting of re-fused glass that appeared due to tumble finishing at elevated temperature (Figure 2a–d). The re-fused glass layer obviously differs in the composition of silicate units from glass of the core. Additionally, its specific volume differs from that of the core: it is larger or smaller than the specific volume of the core depending on the cooling rate of a bead after tumbling. At any rate, an interfacial domain between the core and the shell is always highly stressed in beads that are subject to shell forming (highly likely, in ones made of relatively fusible glass). As a consequence, a fracture eventually arises at the core-shell interface (Yuryeva et al. 2018b), which accelerates the bead fragmentation (Yuryeva et al. 2018a).

Thus far, we have explored the unstable beads mainly using various scanning electron microscope (SEM)-based techniques. Recently, we have presented our Raman study of the 19$^{th}$ century turquoise beads at different stages of degradation and demonstrated substantial changes of the glass structure at the late stages of corrosion, as well as its changes depending on depth in subsurface layers of a degraded glass grain (Balakhnina et al. 2018). Besides, we have determined for the first time the characteristic Raman bands of orthorhombic KSS.

This paper presents the results of the Fourier-transform infrared (FTIR) spectroscopy study of historical turquoise glass beads. Since the vibrational spectroscopy is a powerful method for investigation of glasses, we have focused on the structural characterization of beads at different stages of corrosion (intact, severely cracked with colour alteration, fragmented) and its relationship with the elemental composition.

FTIR spectroscopy is widely used for properly understanding of the changes in physical properties of different types of silicate glass, in particular for lead oxide based glasses since they have been applied in optical and optoelectronic devices. Generally, such research works are devoted to studying of binary silicate glasses (Dalby and King 2006; Piriou and Arashi 1980, Feller et al. 2010)and much less of ternary systems (Mocioiu et al. 2013, Roy 1990). In contrast to predetermined composition of the above-mentioned glasses, the variable content of historical seed beads is very complicated and includes both glass-forming substances (processed vein quartz, or another silicate raw material, which usually have different impurities (see, e.g., [Yasamanov and Yuryev 2001]), pearl ash[1] or potash, calcium carbonate, lead-containing components), and diverse technological additives (fluxes, oxidizing agents, fining agents, colorizing, decolorizing or opacifying agents).

---

[1] An impure product obtained by partial purification of potash from wood ashes.

**Experimental**

*Techniques and equipment*

FTIR spectroscopy analysis was performed using a LUMOS microscope (Bruker) in attenuated total reflection (ATR) mode with Ge ATR crystal. Experiments were carried out in spectral range from 600 to 4000 cm$^{-1}$ at a spectral resolution of 4 cm$^{-1}$ (routine 64 scans were made). The beads were fixed on indium holders with little push (hardness of indium is 1.2 Mohs' scale). This procedure was chosen in view of the complex geometric shape of glass beads and their fragments, fragile condition of samples at some stages of deteriorating, and necessity in further exploration ones by other analytical methods.

FTIR spectra of some glass beads were recorded between 100 and 600 cm$^{-1}$, using a Vertex 70v (Bruker) in ATR mode with diamond ATR crystal. The spectral resolution was 4 cm$^{-1}$, the number of scans was 64.

Elemental compositions of all samples were analysed using a M4 TORNADO micro-X-ray fluorescence (micro-XRF) spectrometer (Bruker). It is equipped with a polycapillary lens, which enables focusing the tube radiation and concentrating it in a spot of about 25 μm in diameter. The X-ray detector is a silicon drift detector with a 30 mm$^2$ active area and energy resolution of about 145 eV. The presence of vacuum sample chamber permits identifying light elements that is very important for glass exploration, as Mg, Al, Si, K, and Ca are significant glass components and influence its properties.

Scanning electron microscopy (SEM) studies were performed using Vega-II XMU (Tescan) and Mira 3 XMU (Tescan) microscopes. As a rule, SEM images are demonstrated in both backscattered electrons (BSE) and secondary electrons (SE) in this article since SEM operating in the former mode represents mainly the substance elemental composition (i.e. its density), while operating in the latter mode it renders mainly the spatial relief, whereas both types of the information are usually required to correctly interpret data obtained using SEM.

*Samples*

Samples of cloudy turquoise glass beads at different stages of destruction were taken during the restoration of secular beaded articles of the 19$^{th}$ century from museum collections of Russia (examples of specimens are shown in Figure 3a). Transparent white and green, and semi-opaque yellow beads were also subjected to study because of differences in their elemental and molecular composition, apart from the turquoise glass ones (Figure 3b).

The surface of dirty glass beads was initially examined to determine the main contaminants, which appeared to be organic compounds and could be the traces of handling, organic pesticides, or organic solvents which were used in previous restoration. Then, before all experiments, samples were washed with high purity isopropyl alcohol ([$C_3H_7OH$] > 99.8 wt. %) at 40°C for 20 minutes in a chemical glass placed into an ultrasonic bath (40 kHz, 120 W). Sample washing was repeated before every experiment.

**Experimental results and interpretation**

*Micro-X-ray fluorescence spectroscopy*

The chemical composition of all historical glass bead samples was studied by micro X-ray fluorescence analysis. As far as XRF is non-destructive and non-invasive method of elemental analysis, it allows one to analyse these samples using other methods (Fusco 2010).

The results of elemental analysis of some samples are presented in the Table 1; they show the distinction in the ratio between the main components and technical additives in the beads of different colours.

These data have shown that lead-potassium glass $PbO–K_2O–SiO_2$ with small content of Cu and Sb was used for turquoise beads manufacture. Investigation of the samples at all stages of destruction indicates that the qualitative elemental composition of glass is the same with small deviations of several elements (e.g., P, S, Ni). Most likely, they are the part of uncontrolled impurities of the main components (for instance, it could be connected with different impurities of the used quartz raw material [Yasamanov 2001] or the elemental composition of pearl ash, which depends on the species of plant [Galibin 2001]). Copper is likely a chromophore of the turquoise glass beads (Sergeyev 1984).

For transparent and green glass beads production, lead-potassium glass was also used, yet the ratios between Pb and K contents of the samples are different. In the transparent glass beads, the contents of Pb and K are the same, and the green ones show a high level of Pb and very low of K. The yellow glass beads are made of lead glass $PbO–SiO_2$ (potassium content is insignificant).

*Micro-FTIR Spectroscopy*

Infrared spectroscopy is widely used in the studies of vitreous and crystalline polymorphous modifications of silicon oxide $SiO_2$. It was determined that all of them are mainly characterized by the presence of two absorption peaks in the region from 600 to 1700 cm$^{-1}$: a strong and wide peak at 1100 cm$^{-1}$ assigned to stretching vibrations of the Si-O bond and a weak wide peak at 800 cm$^{-1}$ assigned to the bending vibrations of the Si–O bond (Lippincott et al. 1958; Moore et al. 2003; Vorsina et al. 2011).

FTIR spectra of the turquoise glass seed beads at different stages of deterioration (Yuryeva and Yuryev 2014; Yuryeva et al. 2017) are presented in Figure 4. A strong and wide band in the region from 820 to 1250 cm$^{-1}$ and weak bands in region from 650 to 820 cm$^{-1}$ are observed in all spectra. Two major absorption bands are seen in the spectra to be shifted to low-frequency range (~1000 and ~775 cm$^{-1}$) compared to ones typical for amorphous $SiO_2$ (~1100 and 800 cm$^{-1}$, respectively). This effect is associated with adding of supplementary components (in this case, lead and potassium), changing the glass structure and, consequently, its infrared spectrum (Roy 1990; Taylor 1990).

Analysis of the FTIR spectra revealed the following features. (i) An appreciable shift of the absorption band 1100 → 1070 cm$^{-1}$ and a less pronounced shift of the absorption band 800 → 775 cm$^{-1}$ is observed in the samples (A1) to (A3); it is also necessary to note the appearance a new peak above 945 cm$^{-1}$. (ii) For the samples (A4) to (A7), the shift of the band at 1100 cm$^{-1}$ is more significant, while the shape of this peak becomes less complicated. (iii) A weak absorption band at 716 cm$^{-1}$ is observed in all spectra. (iv) A weak absorption band above 1630 cm$^{-1}$, which is connected with the presence of water on the surfaces of beads (Gold and Burrows 2004), is also observed. After annealing of the beads at 300°C, the band disappeared completely in less deteriorated samples and became significantly weaker in more destroyed ones. Based on that we conclude that there is adsorbed molecular water in the glass beads surface.

The structure of silicate glasses is known to be dominated by discrete silicate units with short-to-medium range order, which define their physical and chemical characteristics (Dalby and King 2006). In order to determine the structural units in wide absorption bands of complex shapes with several maxima, the deconvolution has been performed using Gaussian functions, and vibration frequencies have been analyzed. The peak positions were found in the second derivative of FTIR spectra. According to the obtained results, the spectra of more corroded samples have fewer fit components than those of the intact ones. For example, five spectral components have been identified for deteriorated turquoise glass bead in the range from 600 to 1300 cm$^{-1}$ (Figure 5a) in comparison with well-preserved samples, which have, as a rule, ten or more fit components (Figure 5b). The deconvolution results for well-preserved turquoise, yellow and green glass seed beads are characterized by more uniform distribution of components and intensity of fit peaks (Figure 5a, c, d). Several strong absorption bands with a close intensity are usually observed in the spectra of these samples. For instance, three intense components with similar integral intensities (17.99, 18.55, and 15.83) are observed at 875, 982 and 1048 cm$^{-1}$, correspondingly, in the green sample (Figure 5d). The bands are assigned to paired tetrahedra ($Si_2O_7^{6-}$), chain ($Si_2O_6^{4-}$) and framework ($SiO_2$) in compliance with the vibration frequencies (Dalby and King 2006). Apart, for strongly deteriorated turquoise beads and transparent glass beads, which also prone to destruction, the main dominant component is observed at ~ 1000 cm$^{-1}$ and ~ 1020 cm$^{-1}$, respectively (Figure 5b, e). Absorptions band at 1000 to 1020 cm$^{-1}$ is characteristic to $SiO_4$ chains when the non-bringing oxygen atoms are bonded to both alkali metal cations ($K^+$) and $Pb^{2+}$ ions (Mocioiu et al. 2013). So, more corroded beads have more depolymerized glass structure in compliance with integral intensity of fit components (chains and paired tetrahedra, mainly).

This result agrees well with the Raman spectroscopy data. The appearance of an intense line of Raman scattering peaked at 980 cm$^{-1}$ (Balakhnina et al. 2018), usually associated with the $Si_2O_6^{4-}$ units, in the degraded turquoise bead glass evidences in support of the simplification of the glass structure in the course of its degradation and dominance of the silicate chain units in damaged glass.

The obtained data also correlate with the results of the FTIR study of turquoise glass beads in the far spectral region. The FTIR spectrum of turquoise glass bead at the intermediate stage of destruction (A3) in the region from 100 to 600 cm$^{-1}$ is presented in Figure 5f. The spectrum shows two absorption bands at 286 and 245 cm$^{-1}$ assigned to Pb–O–Pb bonds in the [$PbO_4$] unit that confirm the presence of chains (Mocioiu et al. 2006). Lead may act in glasses as a network former or as modifier of the $SiO_2$ network. The absorption band at 431 cm$^{-1}$ is assigned to Pb–O–Si bonds that prove the presence of the [Pb$_2$O$_4$] units in bead glass of this type. In the sample A3, integral intensity of fit components assigned to sheets and framework ([Pb$_2$O$_4$] units) have greater contribution compared to more deteriorated samples A6 and A7.

**Brief Discussion**

The observed simplification of the glass structure during its corrosion may reduce the glass strength and, hence, accelerate the generation of micro-discontinuities and microcracks in its bulk due to internal stress inherent to bead glasses prone to degradation (Yuryeva et al. 2018a); additionally, depolymerization facilitates the growth of fractures. It was shown by Pukh et al. (2005) that the ultimate elastic strain experienced by glass at the instant of fracture is about 10 % for glasses with a three-dimensional atomic structure and about 5 % for glasses with a layered or chain structure. This fact is associated in that article with the decreased uniformity of the loading distribution over atomic bonds in less polymerized glasses. Thus, following Pukh et al. (2005), we have come to conclusion that the observed glass depolymerization is another reason for the bead glass corrosion.

The phenomenon of glass depolymerization may be understood taking into account the fluctuation mechanism of the fracture at atomic level (Slutsker 2005). The process resulting in fracture of a stressed solid is known to comprise a sequence of elementary events of stressed atomic bonds breaking by local energy fluctuations. It may be assumed that energy fluctuations in internally stressed glass result in, foremost, reduction of atomic bond breaking barrier and accelerated rupturing of bonds forming first three-dimensional and then two-dimensional silicate networks likely due to excess stress applied to them because of non-uniform distribution of the local loading among the bonds. Chains and especially paired tetrahedra appear to be more resistant to stress, presumably because of lower loading applied to bridging oxygen bonds in them and/or due to higher initial potential barrier of breaking these bonds. Thus, in such a scenario, internal stress induced glass depolymerization, decreasing glass strength, should be considered as an important degradation mechanism inherent to this type of glass corrosion.

It should be noted also that the characteristic dimensions of light scatterers (estimated as *a* << 150 nm) observed by us in deteriorating turquoise beads glass (Yuryeva and Yuryev 2014), the number density of which grows with the rising glass corrosion, coincide with the sizes of nucleating micro-discontinuities that are typical defects emerging at locations of shear and rotational strains (Betekhtin and Kadomtsev 2005). We presumably identify the observed light scatterers with this type of defects. The material destruction related to formation and growth of such microscopic defects is a kinetic thermal fluctuation process occurring throughout the entire period of material loading (Betekhtin and Kadomtsev 2005); their merger and formation of macrocracks is the only possible final phase of this process.

**Conclusion**

In the work, we make an effort to establish correlation between turquoise glass beads preservation, their elemental composition and changes of glass structure. For comparison, prone to destruction transparent white glass beads and well-preserved stable green and yellow glass beads are also objects of study.

Micro-XRF spectrometry of samples has shown that lead-potassium $PbO–K_2O–SiO_2$ glass with a small amount of Cu and Sb was used for turquoise beads manufacture. In addition to that, many uncontrolled impurities, which could be connected with different admixtures of main components and technical additives, were detected. FTIR study shows that more corroded beads have more depolymerized structure in compliance with integral intensity of fit components (chains and paired tetrahedra, mainly). Such simplification of the glass structure may reduce the glass strength and promote the formation of new microcracks in its volume as well as the growth of existing ones.

Finally, we can conclude that the model of bead glass internal corrosion, previously proposed by us for explanation of the long-term process of the turquoise bead glass degradation at room temperature (Yuryeva et al. 2018b), agrees with the commonly adopted thermal fluctuation theory of materials fracture. The simplification of the glass structure under internal stress during the long-term degradation of glass at room temperature also may find a realistic explanation within this theory.

We consider glass depolymerization, caused by the internal stress and decreasing the glass strength, as an essential corrosion mechanism of stressed glass of seed beads.

**Acknowledgments**

This work was supported by the Russian Science Foundation under Grant number 16-18-10366. We thank our research team members Mr. Ilya Afanasyev of The Russian Federal Center of Forensic Science of the Ministry of Justice for SEM images and Miss Darya Klyuchnikova of GosNIIR for preparation of samples to experiments. We also grateful to Ms. Lyubov Pelgunova of A. N. Severtsov Institute of Ecology and Evolution of RAS for XRF analyses.

**Conflict of interest**

No potential conflict of interest was reported by the authors.

https://arxiv.org/abs/1705.09394

https://arxiv.org/abs/1705.09394

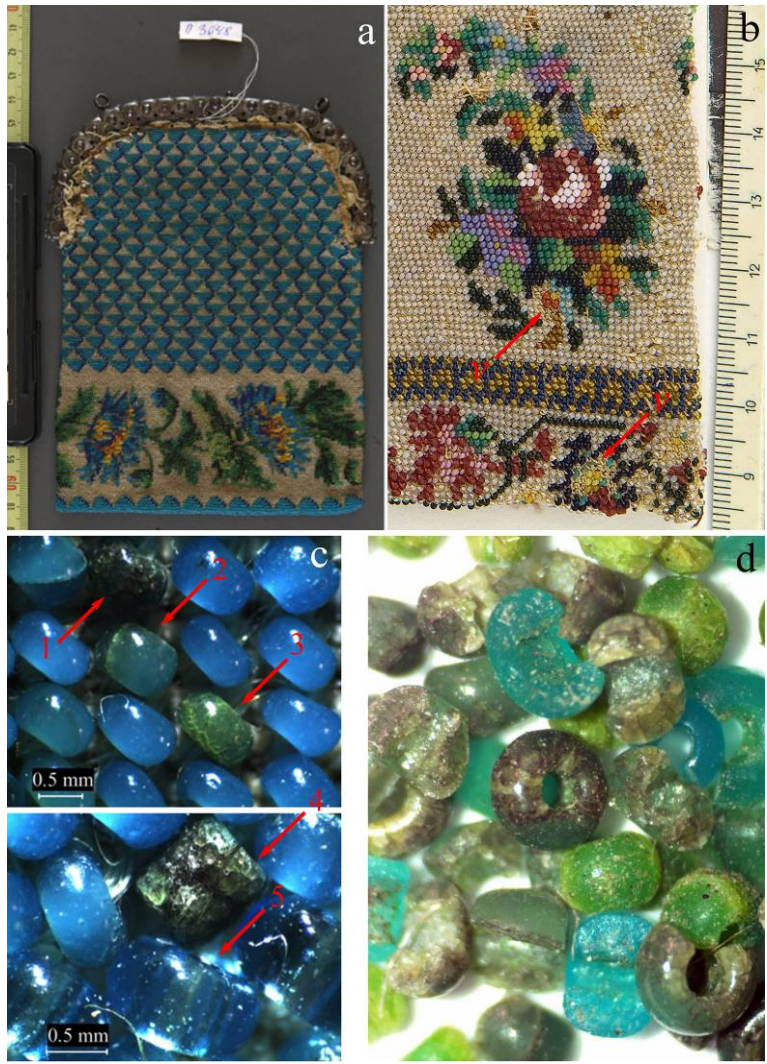

Figure 1. (a) A view of a beaded reticule and (b) an area on another beaded reticule of the 19[th] century, collection of the Museum of A. S. Pushkin in Moscow; both photographs were taken before restoration: arrows marked with the letters '*v*' show the voids where turquoise glass beads have been lost; at the same time, the areas embroidered with beads of other colours are in fairly good condition, and the beads themselves remain transparent and well-preserved; (c) a photograph of damaged beadwork on the reticule shown in the panel (a): arrows numbered from 1 to 4 show examples of historical turquoise beads at different stages of destruction surrounded by the intact ones; the arrow 5 indicates a modern blue bead replacing the lost historical turquoise one (the former is larger and more transparent); (d) corroded (cracked, discolored and crumbled) and mechanically damaged turquoise seed beads from a beaded museum exhibit of the 19[th] century obtained during its restoration (the beads are shown unwashed).

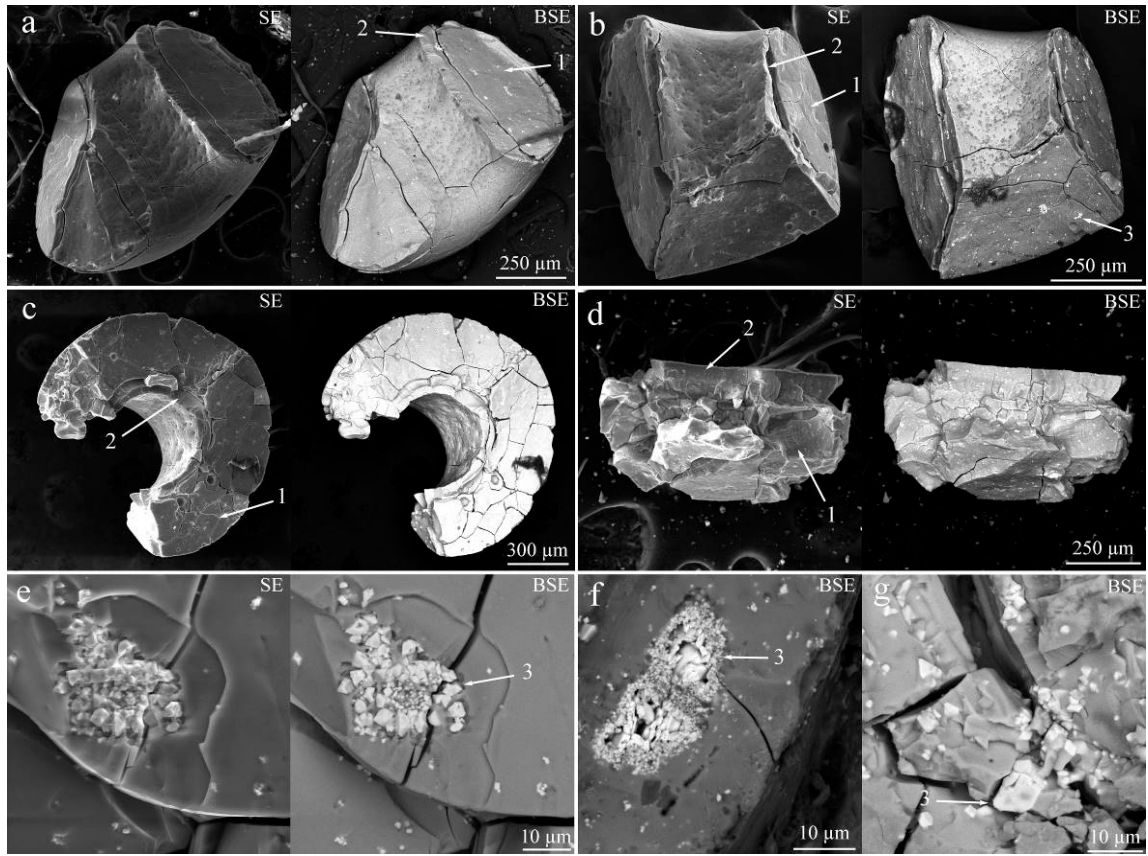

Figure 2. SEM SE and BSE images of several turquoise bead glass samples demonstrating domains of original and re-fused glass—cores (1) and shells (2)—and colonies of $KSbOSiO_4$ (KSS) crystallites in glass (3): (a) and (b) segments of fragmented beads from the same exhibit; (c) an unusually fragmented bead showing details of its interior region; (d) sand-like particles of a bead after complete breaking; (e) to (g) KSS colonies causing the internal stress and glass fracture.

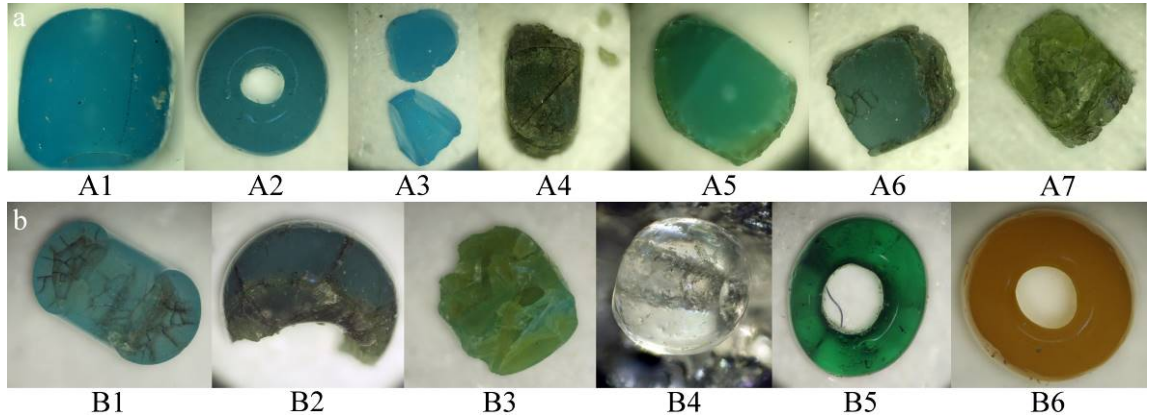

Figure 3. (a) Turquoise glass beads at different stages of degradation: (A1), (A2) the appearance of the first cracks on the surface of intact blue beads, (A3) splinters of beads, (A4) to (A7) fragments of beads with corrosion marks: significant color changes, an increased number of cracks and brown domains around them; the corresponding FTIR spectra are presented in Figure 4; (b) photographs of the samples of turquoise (B1) to (B3), transparent (B4), green (B5), and yellow (B6) glass seed beads examined using micro-XRF spectrometry; the elemental composition data for these samples are given in Table 1.

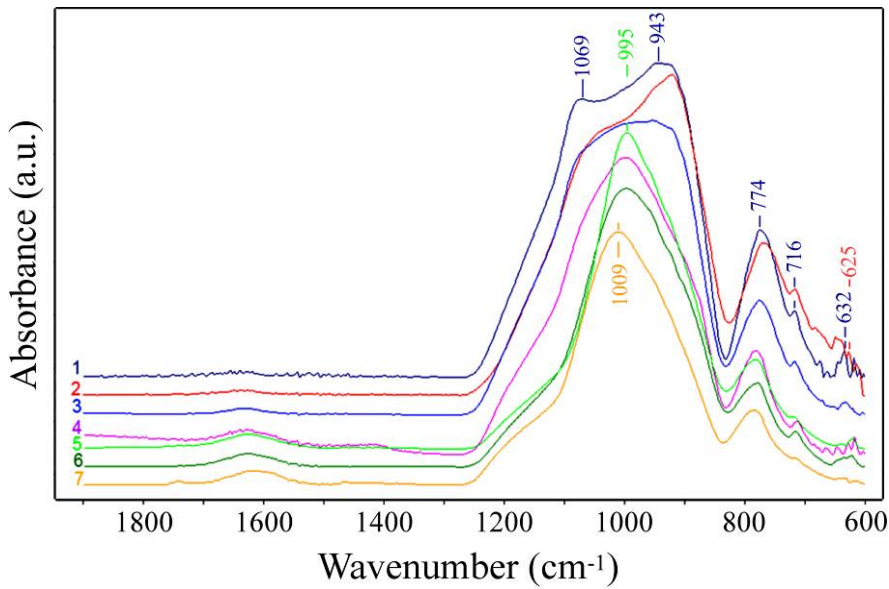

Figure 4. FTIR absorption spectra for the samples of turquoise bead glass at different stages of corrosion: curve numbering corresponds to sample numbering in Figure 3a (A1 to A7, respectively).

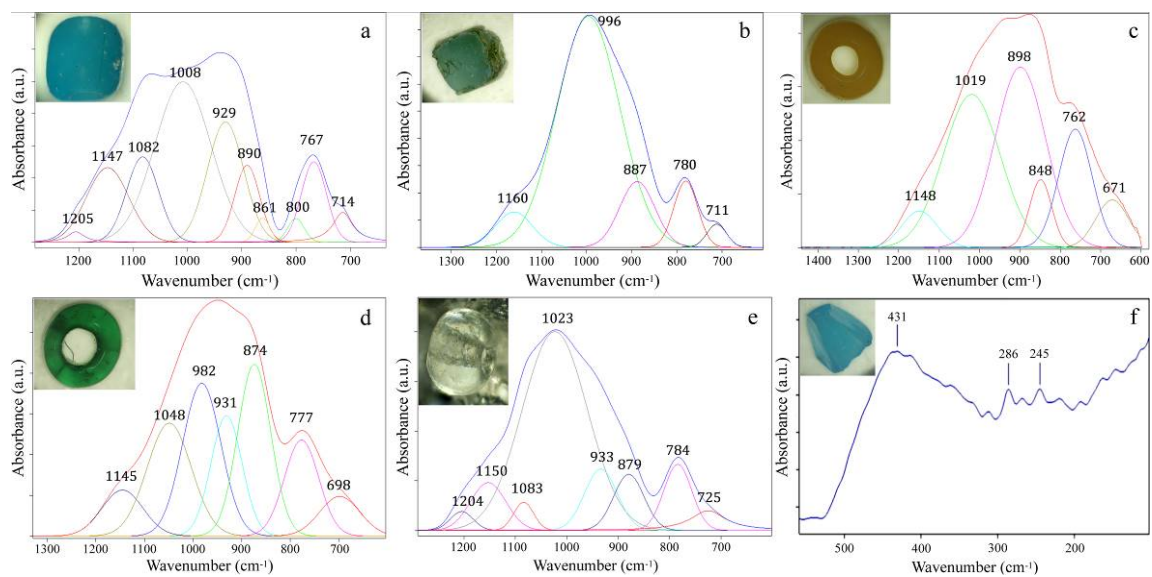

Figure 5. FTIR absorption spectra for the samples of different bead glass (the sample photos are shown in the inserts and in Figure 3): (a) an intact turquoise bead (sample A1), (b) a piece of highly corroded turquoise bead glass (sample A6), (c) an intact (stable) yellow bead (sample B6), (d) an intact (stable) green bead (sample B5), (e) a transparent white bead (prone to degradation, sample B4), (f) a piece of turquoise bead glass at the intermediate stage of corrosion (sample A3).

Table 1. The results (at. %) of elemental analysis of samples of turquoise (B1–B3), transparent (B4), green (B5), and yellow (B6) glass seed beads (Figure 3b).

|    | Mg   | Al   | Si    | P    | S    | K     | Ca   | Mn   | Fe   | Ni   | Cu    | Zn   | As   | Sb   | Pb    |
|----|------|------|-------|------|------|-------|------|------|------|------|-------|------|------|------|-------|
| A1 | -    | -    | 38.41 | -    | 0.91 | 19.07 | 0.98 | -    | 0.21 | -    | 3.71  | 0.06 | 0.28 | 5.49 | 30.88 |
| A2 | -    | 0.79 | 45.21 | -    | -    | 16.41 | 1.95 | 0.02 | 0.32 | -    | 6.48  | 0.08 | 0.50 | 5.37 | 22.85 |
| A3 | -    | 0.67 | 46.60 | -    | 0.09 | 12.50 | 2.63 | 0.07 | 0.38 | -    | 6.51  | 0.07 | 0.53 | 5.09 | 24.86 |
| A4 | -    | 0.66 | 43.60 | 0.05 | 0.19 | 17.50 | 1.14 | -    | 0.29 | -    | 10.83 | 0.10 | 0.57 | 6.03 | 19.04 |
| A5 | -    | 0.77 | 47.69 | -    | 0.06 | 10.80 | 0.94 | 0.01 | 0.27 | 0.03 | 8.75  | 0.10 | 0.37 | 7.31 | 22.75 |
| A6 | -    | 0.69 | 45.07 | -    | -    | 18.91 | 2.13 | 0.04 | 0.68 | -    | 6.75  | -    | 0.69 | 6.14 | 18.90 |
| A7 | -    | 0.66 | 43.60 | 0.05 | 0.19 | 17.50 | 1.14 | -    | 0.29 | -    | 10.83 | 0.10 | 0.57 | 6.03 | 19.04 |
| B1 | -    | 1.11 | 54.31 | -    | -    | 30.25 | 6.33 | 0.02 | 0.30 | -    | 1.84  | 0.03 | 0.13 | 2.16 | 3.52  |
| B2 | -    | 1.04 | 68.80 | -    | 0.71 | 16.19 | 0.59 | 0.01 | 0.19 | 0.02 | 6.29  | 0.08 | 0.30 | 1.57 | 4.22  |
| B3 | -    | 1.02 | 65.03 | 0.23 | 0.24 | 18.70 | 1.19 | -    | 0.21 | -    | 7.13  | 0.06 | 0.31 | 2.07 | 3.85  |
| B4 | 1.91 | 1.08 | 71.04 | -    | -    | 14.08 | 1.27 | 0.1  | 0.12 | -    | 0.03  | 0.02 | 1.04 | -    | 9.31  |
| B5 | -    | 1.35 | 56.31 | -    | 2.17 | 2.07  | 5.41 | 0.79 | 1.01 | 0.02 | 3.40  | 0.07 |      | -    | 27.4  |
| B6 | -    | 1.54 | 52.72 | 0.07 | 7.03 | 0.22  | 2.37 | 0.1  | 0.71 | -    | 0.2   | 0.04 | 0.54 | 0.98 | 33.48 |